# Impact of atmospheric impairments on mmWave based outdoor communication

Pavithra Nagaraj
*PG Scholar, Digital Communication Engineering,*
*Department of Electronics and Communication Engineering, BMS College of Engineering, India*

**Abstract**

Substantial growth of demand for mobile data and rapidly increasing spread of personal communication devices such as smart phones and tablets is the major challenge faced by the mobile service providers in recent years. The allotted spectrum for the currently operating mobile communication systems have been saturated in the last few years by such a considerable rate that the future's fifth generation network would not be able to operate without applying new frequencies. In accordance with the concept of 5G, millimeter wave frequency band will also be used along with the currently used frequency bands. In this particular spectrum however, a new and most significant attenuation factor due to precipitation and atmospheric absorption is introduced. In this paper the effects of precipitation, fog and atmospheric absorption on millimeter wave propagation are investigated using MATLAB.

## Introduction

With the substantial growth of demand for mobile traffic, the trade-off between capacity requirement and spectrum shortage becomes increasingly prominent. Current mobile communication services (3G and 4G as well) use carrier frequencies ranging from 700MHz to 2.6GHz for mobile data traffic. The several different communication systems however, cover this spectrum with disjoint frequency bands leaving narrow pieces of unused frequencies scattered in the 2GHz range [3].

The bottleneck of wireless bandwidth becomes a key problem of the next generation wireless networks [4]. The tremendous amount of raw bandwidth at mmWave frequency band can deliver multiple Gigabits per second (Gbps) data rates and can relieve the mobile data traffic congestion in lower frequency bands [5]. 5G cellular systems are likely to operate in or around the millimeter wave (mmWave) frequency band of 30-300GHz as shown in Fig.1, where vast spectrum currently exists in light use [6]. Millimeter wave frequencies have much smaller wavelengths, ranging from 1mm to 10mm, almost comparable to the size of a human finger nail, whereas 4G frequencies have wavelengths that are tens of centimeters. Smaller wavelengths at these frequencies have often been thought to result in higher attenuation (due to the precipitation and oxygen absorption) through air, than that observed at today's cellular bands. Propagation loss is more in mmWave band compared to the low microwave regime.

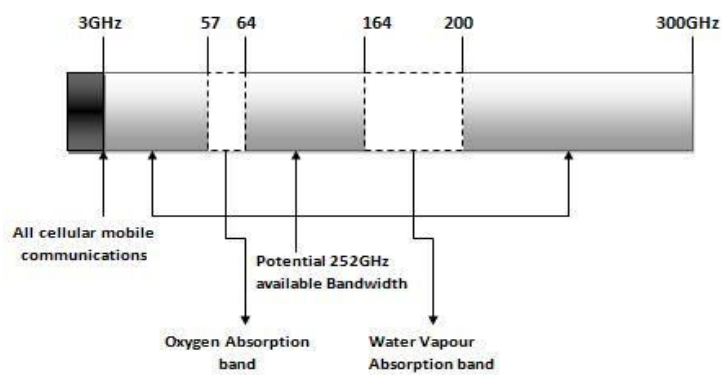

Fig.1. 3GHz-300GHz mm-Wave spectrum

Until recently, mmWave frequencies spanning from 30GHz to 300GHz were not considered useful in dynamic communication environments such as cellular systems. Millimeter waves have been used broadly for long distance point-to-point communication in satellite and terrestrial applications [7], now they are being investigated and developed for commercial cellular systems. This novel application is much more challenging due to the unpredictable propagation environments and strict constraints on the attributes like cost, size, and power consumption (particularly in the mobile handset).

The propagation properties for mmWave band are widely different from those of lower frequency bands. Rain induced signal attenuation is most significant for terrestrial microwave links which operate at frequencies higher than 10GHz. Rain induced attenuation severely degrades the radio wave propagation at centimeter or millimeter wavelengths due to absorption and scattering of radio waves, which result in reduction of the received signal level, further restricts the range of radio communication systems and hence limits the use of higher frequencies for line-of-sight microwave links. In addition to the rain attenuation, attenuation due to atmospheric gases, scintillation and attenuation due to fog are most significant factors which affect the use of millimeter wave frequencies. So, in order to aid the development of mmWave systems, the propagation characteristics of mmWave wireless channels need to be characterized, modelled and evaluated by dedicated research.

In this project, we intend to investigate the effects of the propagation factors such as attenuation by rain and fog, scintillation, attenuation due to water vapour and oxygen absorption on mmWave frequencies. Suitable frequencies for indoor and outdoor communication are estimated depending on the attenuation levels.

## Propagation factors

The characteristics of millimeter wave communications should be considered in the design of architecture and protocols to fully exploit their potential. In the following subsections we present and summarize the characteristics and propagation factors of millimeter frequency band.

### *Free space pathloss*

The pathloss can be defined as the ratio of the transmit power to the received signal power as a function of the Tx-Rx separation distance. A good pathloss model is essential for the analysis of link budget and network planning. The free space pathloss (FSPL), which is the pathloss for two isotropic antennas placed in free space separated by a distance d, is given by

$$\text{FSPL (in dB)} = \left(\frac{4\pi d}{\lambda}\right)^2 = \left(\frac{4\pi d f}{c}\right)^2$$

Here, it is seen that the free space path loss is proportional to square of the carrier frequency, i.e. FSPL $\propto f^2$, resulting in a much more severe FSPL for mmWave systems.

For typical radio applications, the frequency f is commonly measured interms of GHz and distance d is measured in terms of Km, in this case the free space propagation loss equation becomes

$$\text{FSPL (in dB)} = 20 \log_{10}(d) + 20 \log_{10}(f) + 92.45$$

Fig.2 shows the free space pathloss at 2.4, 28 and 100 GHz, as function of Tx-Rx separation distance.

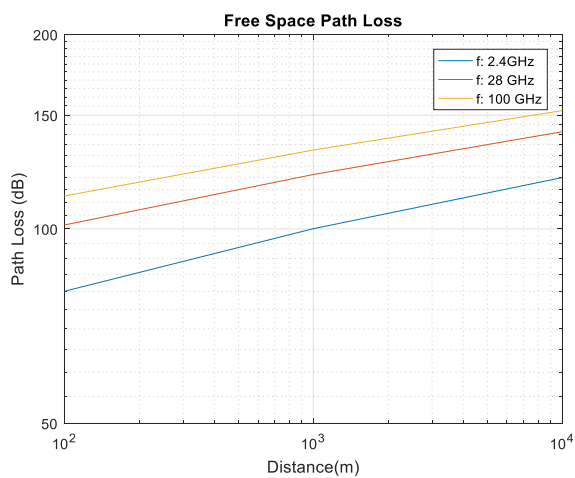
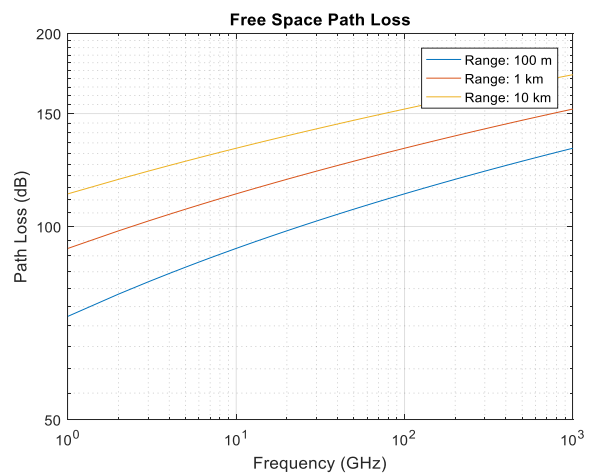

Fig.2. Free space pathloss at 2.4, 28 and 100 GHz as function of distance.

Fig.3. Free space pathloss at T-R distance of 100m, 1Km and 10Km as function of frequency

Compared to the free space pathloss for a given distance at 2.4GHz, the FSPL is about 22dB larger for 28GHz and about 32dB larger for 100GHz.

Fig.3 shows the free space pathloss at Tx-Rx separation distance (Range) of 100m, 1Km and 10Km, as function of frequency.

Compared to the free space pathloss for a given frequency at 100m separation distance, the FSPL is about 20dB larger for 1Km and about 42dB larger for 10Km. Both the Fig.2 and Fig.3 illustrates that the pathloss increases with range and frequency.

## Propagation Loss due to Rain

The free space pathloss describes only a part of signal attenuation which occurs while travelling in vacuum. But in reality, signals do not travel in vacuum; Signals interact with particles in the air and lose energy as they travel along the propagation path.The attenuation varies with different factors such as temperature, water density and pressure.

Rain is usually characterized by the rain rate (in mm/hr) .The rain rates for different weather conditions are listed below,

Very light rain (Drizzle) < 0.25mm/hr,
Light rain – 0.25 mm/hr to 1 mm/hr,
Moderate rain – 1mm/hr to 4 mm/hr,
Heavy rain – 4 mm/hr to 16 mm/hr,
Extreme rain – 16 mm/hr to 50 mm/hr,
Torrential rain > 50mm/hr.

In addition, rain attenuation is also a function of the shape of the raindrop, its relative size compared to the RF signal wavelength and the signal polarization.

Fig.4 shows the Rain attenuation in dB/Km across the frequency band at different rainfall rates as per the ITU recommended model [10].

In general, horizontal polarization represents the worst case for propagation loss due to rain. Since we are considered the worst case scenario to investigate the attenuation due to rain, the polarisation is assumed to be horizontal and hence the tilt angle is zero. In addition, assuming that the signal propagates parallel to the ground, the elevation angle is zero. The range of 1Km is considered.

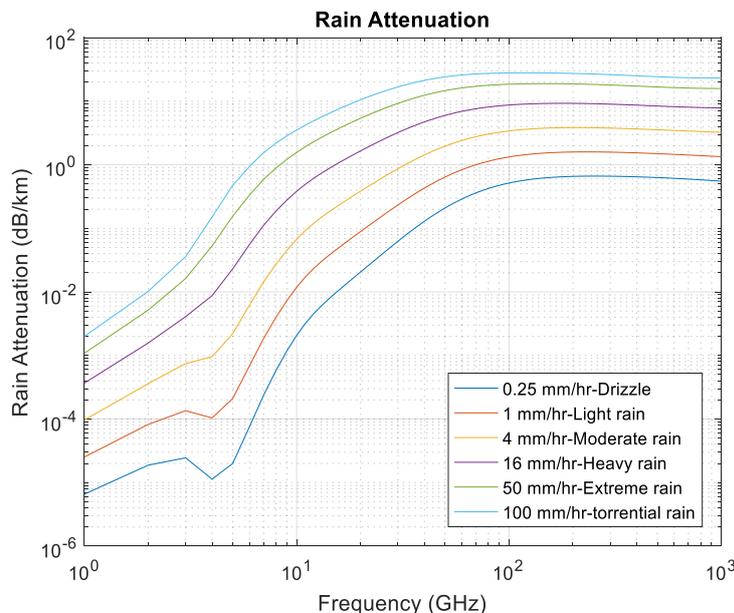

Fig.4. Rain attenuation in dB/Km across the frequency band at different rainfall rates

From Fig.4 we can interpret that rain attenuation is more at frequencies above 100GHz. So, the higher mmWave frequencies are not suitable for outdoor communication. As the rainfall rate increases, the attenuation level also increases. Hence, rain could be a major limiting factor for mmWave communications, especially when operating above 10GHz. We can observe that a heavy rainfall can cause around 5dB/km attenuation at 28GHz.

## Propagation Loss due to Fog

Cloud and fog are also structured by water droplets but of much smaller size compared to that of rain drops. The size of the fog droplets are generally lesser than 0.01cm. Fog is usually characterized by two parameters-liquid water density and atmospheric temperature.

A medium fog with visibility of approximately 300 meters, has a liquid water density of 0.05 g/m^3.For heavy fog where the visibility drops to 50 meters, liquid water density is around 0.5 g/m^3. The atmospheric temperature (in degree Celsius) is also present in the ITU recommended model for propagation losses due to fog and cloud [11]. Pathloss due to cloud and fog are usually minor, but may be crucial in a few rare cases.

The following Fig.5 illustrates how the attenuation due to fog varies with frequency. The standard atmospheric temperature is taken as 15 degree Celsius as per the ITU model. As the level of liquid water density tends to increase, the attenuation also increases. From the figure we can infer that the attenuation caused due to clouds and fog may be a factor of importance especially for microwave systems well above 10GHz.

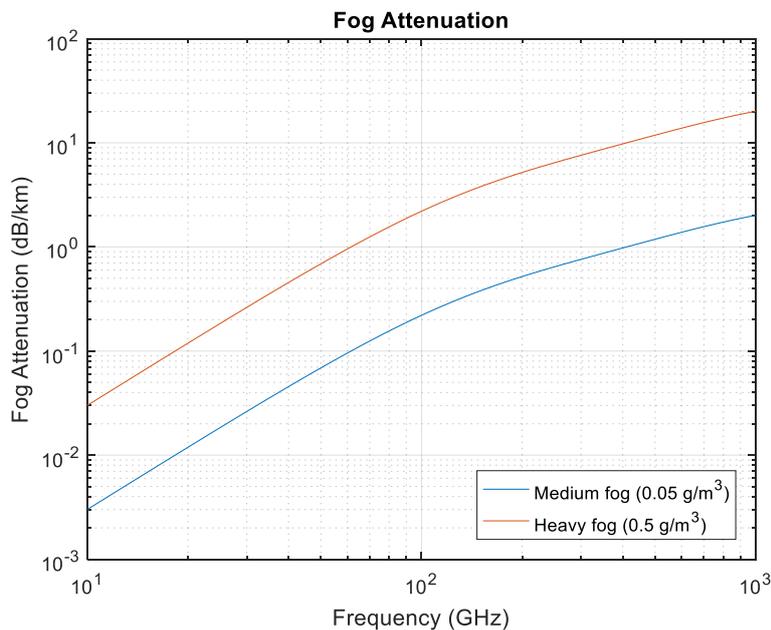

Fig.5. Fog attenuation in dB/Km across the frequency band at different liquid water densities.

*Propagation Loss due to Atmospheric Gases*

Even in the absence of rain and fog, the atmosphere is full of gases that still degrade the signal propagation along the propagation path. The ITU recommended model for attenuation due to atmospheric gases, describes atmospheric gas attenuation as a function of dry air pressure, like oxygen, measured in hPa, and water vapour density, measured in g/m^3[12]. The standard atmospheric pressure is 101325Pa at a standard temperature of 15 degree Celsius.

Fig. 6 shows atmospheric attenuation at different frequencies assuming a dry air pressure of 1013.25 hPa at 15 degrees Celsius, and a water vapour density of 7.5 g/m^3.

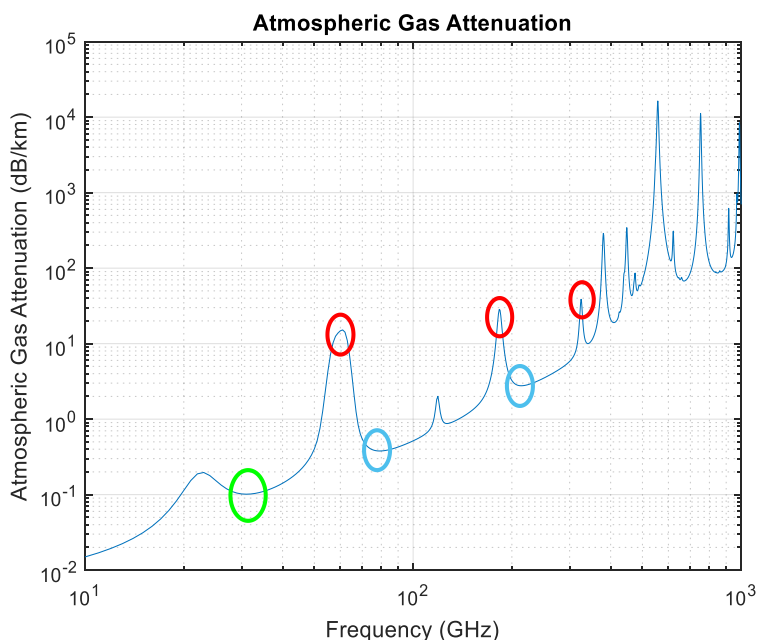

Fig.6. Atmospheric gas attenuation at different frequency bands.

The green circle shows that the frequencies ranging from 28 and 38GHz encounter very small attenuation due to gas, providing feasibility of mmWave communication at such frequencies. It can be observed that negligible atmospheric absorption is encountered at 28GHz and 38GHz, as well as in 70-90 GHz, 120-170GHz and 200-280GHz frequency bands. Hence, the mmWave bands promise a massive amount of unlicensed spectrum from 28GHz to 38GHz, which are the most potential frequency bands for the future 5G cellular systems. The blue circles show the attenuation levels similar to attenuation in current communication systems, comparably larger than the green circle. The red circles indicate frequencies with high attenuation, thus viable for indoor communication.

## Conclusion

This paper has studied the impact of atmospheric impairments on mmWave communication in outdoor environment. The feasibility of utilizing the mmWave spectrum by analyzing different propagation factors to be considered while designing the mmWave outdoor model is analysed .In general the notion of using mmWave for cellular communication is still in its early stages with more work to be done in near future.